\documentclass[acmtocl,acmnow]{acmtrans2m}
\renewcommand{\textwidth}{450pt}
\renewcommand{\textheight}{580pt}
\def\nsubseteq{{\rm /\!\!\!\!\!\!\subseteq}}

\def\N{\mathbf{N}}
\def\Z{\mathbf{Z}}

\def\bx{~~~$\Box$}

\def\N{\mbox{$\sf I\!N$}}

\def\Z{\mbox{$\sf Z\hskip-5.0pt Z$}}
\def\nmid{\mathrel{\kern2.9375pt\mid\kern2.9375pt}}

\def\pf{\noindent {\bf Proof}~~}
\newtheorem{theorem}{Theorem}[section]

\newcommand{\be}{\begin{enumerate}}
\newcommand{\ee}{\end{enumerate}}
\newcommand{\bt}{\begin{theorem}}
\newcommand{\et}{\end{theorem}}
\newtheorem{lemma}[theorem]{Lemma}
\newcommand{\bl}{\begin{lemma}}
\newcommand{\el}{\end{lemma}}
\newtheorem{prop}[theorem]{Proposition}
\newcommand{\bp}{\begin{prop}}
\newcommand{\ep}{\end{prop}}
\newtheorem{defn}[theorem]{Definition}
\newcommand{\bd}{\begin{defn}}
\newcommand{\ed}{\end{defn}}
\newtheorem{remarks}[theorem]{Remarks}
\newtheorem{remark}[theorem]{Remark}
\newcommand{\brem}{\begin{remark}}
\newcommand{\erem}{\end{remark}}
\newtheorem{exercise}[theorem]{Exercise}
\newcommand{\bxr}{\begin{exercise}}
\newcommand{\exr}{\end{exercise}}
\newtheorem{example}[theorem]{Example}
\newcommand{\bxm}{\begin{example}}
\newcommand{\exm}{\end{example}}
\newcommand{\beqa}{\begin{eqnarray*}}
\newcommand{\eeqa}{\end{eqnarray*}}
\newcommand{\bc}{\begin{center}}
\newcommand{\ec}{\end{center}}
\newcommand{\bcent}{\begin{center}}
\newcommand{\ecent}{\end{center}}

\newtheorem{corollary}[theorem]{Corollary}
\newcommand{\bcor}{\begin{corollary}}
\newcommand{\ecor}{\end{corollary}}
\newtheorem{defns}[theorem]{Definitions}
\newcommand{\bds}{\begin{defns}}
\newcommand{\eds}{\end{defns}}
\newcommand{\brems}{\begin{remarks}}
\newcommand{\erems}{\end{remarks}}
\newtheorem{exercises}[theorem]{Exercises}
\newcommand{\bxrs}{\begin{exercises}}
\newcommand{\exrs}{\end{exercises}}
\newtheorem{examples}[theorem]{Examples}
\newcommand{\bxms}{\begin{examples}}
\newcommand{\exms}{\end{examples}}

\def\sP{\mbox{${\cal P}$}}

\def\to{\rightarrow}

\def\To{\Rightarrow}

\def\yields{\mapsto^{*}}

\def\Di{\Delta^{\infty}}
\def\emp{\emptyset}
\title{On the Definition of Nondeterministic Mechanisms}
\author{K.Venkata Rao\\ Government College (A)\\Rajahmundry, A.P., India \and Kasturi Viswanath\\
International Institute of Information Technology, Hyderabad, India}
\date{April 9, 2009}

\begin{abstract}
We present here three different approaches to the problem of
modeling mathematically the concept of a nondeterministic mechanism.
Each of these three approaches leads to a mathematical definition.
We then show that all the three mathematical concepts are equivalent
to one another. This insight gives us the option of approaching
 the $wp$ formalism of Dijkstra from a different viewpoint that
 is easier to understand and to teach.
\end{abstract}

\category{F.1.2 --- Modes of Computation --- Alternation and
nondeterminism}{F.0 --- Miscellaneous}{...}

\terms{Algorithms}


\begin{document}

\begin{bottomstuff}

\end{bottomstuff}

\maketitle

\section{Introduction} \label{sec-intro}

 In his well-known book \cite{Dijk76} Dijkstra speaks of his intention to
present ``a number of beautiful algorithms in such a way that the
reader (can) appreciate their beauty" and do so ``by describing the
... design process that would each time lead to the program
concerned".

He then introduces the {\em wp} formalism. In his hands this
becomes a powerful tool to carry out his agenda. Surely this
methodology should be more widely taught and learned. Not
only that, it is necessary to examine if it can be extended
to cover the present programming paradigms. However, the
$wp$ formalism is hard to learn and use.  One is therefore
interested in exploring alternative approaches to the
formalism that make it simpler to understand and easier to
practise. In this article we show that the backward mapping
predicate transformers that Dijkstra uses may be effectively
replaced by forward mapping state choice maps. It becomes
possible then to use the alternative approach suggested by
the results of this paper to carry out his agenda in a
different and perhaps more transparent manner.

\section{Three Definitions}\label{sec-defns} The mapcode approach to the understanding of computing
concepts in the deterministic case has been elaborated in
\cite{DEDS,KVbook} and in the references cited there.  This approach
models a program as the repeated application of a self-map on a set,
following \cite{Knuth}, page 7. It has been shown \cite{KVbook} that
this generic model is sufficient to convey an understanding of many
concepts ranging from machine language to neural networks. At the
same time it is sufficiently practical to formulate many standard
programs rigorously and prove their total correctness.

It is necessary to extend this approach to the study of parallelism
and concurrency. To this end it is necessary to first choose a
mathematical model for nondeterministic programs.  The mapcode
philosophy suggests that we set aside for the time being the formal
language problems of how to get a machine to do what we want it to
do, and strive for clarity in the language of sets and maps as to
what exactly we want the machine to do and why.

We start with the concept of a {\em state space} $X$. This is the
space of variables on which the computation takes place. As in the
deterministic case we take the point of view that a generic
nondeterministic program consists of repeatedly invoking a
nondeterministic mechanism till a stopping condition is met. Thus
the focus shifts to the modeling of a nondeterministic mechanism.
Looking at the question $ab~initio$ we show that there are three
natural viewpoints. Fortunately all three turn out to lead to
equivalent mathematical structures. We are thus enabled to proceed
with the theory of nondeterministic computation in subsequent
articles basing ourselves on any one of the definitions studied
here.

The first approach is the simplest and the most natural. Given $x
\in X$, let $\Delta(x)$ denote a subset of $X$. We can model
nondeterminism by requiring that if the current state is $x$, then
the mechanism when invoked presents us with one of the states $y$ in
$\Delta(x)$ in a finite amount of time. How exactly the state $y$ is
produced is hidden from us. It is observed in \cite{WM97} that this
is the most common approach.

Our second approach is the one suggested by Dijkstra \cite{Dijk76}.
In this approach the focus shifts from individual states to sets of
states and from initial states to final outcomes. We ask the
question: given a set $A \subseteq X$ what is the set of all states
$\mu(A)$ for which if the initial state $x \in \mu(A)$, then the
mechanism when invoked returns with certainty an outcome that is in
$A$? If we knew $\mu(A)$ for every $A$, then it is reasonable to
feel that we have understood the mechanism well.\footnote{The symbol
$\mu$ has been chosen to represent a multiplicative map. Later on,
we use the symbol $\alpha$ to denote an additive map.}

 Because the map $\mu : \sP(X) \to \sP(X)$ has been derived by a particular line of
reasoning, it is automatically endowed with certain properties. For
example, if $A \subseteq B$, then we should have $\mu(A) \subseteq
\mu(B)$. After all, if starting in $\mu(A)$ guarantees that we will
move into $A$, it should also guarantee that we will move into $B$,
because $A \subseteq B$. It is also reasonable to require that $\mu$
should carry the empty set to the empty set.

Let $\{A_j~|~j \in J\}$ be any collection of sets.  If starting in
$\mu(A_j)$ guarantees the outcome to be in $A_j$, then staring in
$\cap \mu(A_j)$ should guarantee  the outcome to be in $\cap A_j$.
So it is necessary that $\cap \mu(A_j) \subseteq \mu(\cap A_j)$.
Because $\cap A_j \subseteq A_j$ for all $j$, by the monotonicity
property of $\mu$ just observed, the reverse inequality also holds.
So we must have $\mu (\cap A_j ) =
 \cap \mu(A_j)$.

In the case of unions the monotonicity property implies that $\cup
\mu(A_j) \subseteq \mu( \cup A_j)$. A little reflection shows that
the reverse inequality need not hold. Given a state $x$ one may be
able to guarantee that the outcome $y$ is in the union, though $y$
may not be uniquely determined by $x$. It is possible that  $y$
could belong to one $A_j$ on one invocation of the mechanism and in
another $A_j$ for another invocation. So we may not be in a position
to say that the outcome $y$ will definitely be in one of them.

In the discussion above, we used the monotonicity property to
establish the intersection preserving property. It is possible to
show, and we shall do so later, that the intersection preserving
property implies the monotonicity property. So let us choose the
defining properties  of $\mu$ to be $\mu(\emp) = \emp$ and $\mu
(\cap A_j ) = \cap \mu(A_j)$. $\mu$  is our second definition for a
nondeterministic mechanism.

The third approach is similar to the second. Now we ask:
given a
 set $A$  what is the set $\alpha(A)$
of all states with the property that we can guarantee that at least
one of the outcomes will be in $A$? (Earlier we wanted every
possible outcome to be in $A$, now we only ask for at least one
outcome in $A$.) Repeating the thought processes that led us to
derive the properties of $\mu$ it is not difficult to conclude that
$\alpha$ should carry the empty set to itself and preserve arbitrary
unions.

In the rest of the article we shall study these three
definitions mathematically  and show how they relate to one
another and to Dijkstra's  theory.

\section{Choice Set Maps}

In what follows $X$ denotes a  non-empty set called the {\em
state space}. $\sP(X)$ is the powerset of $X$. $\emp$
denotes the empty set. $A$ will denote an arbitrary subset
of $X$ and $\{A_j\}$ will be an arbitrary collection of
subsets of $X$. The symbol $\doteq$ may be read as `is
defined to be'.

 \bd {\rm    A map $\Delta : X \to \sP(X)$ is called
a {\em choice set map} on $X$. $\Delta(x)$ is called the
{\em choice set} at $x$. The pair $(X,\Delta)$ is called a
{\em choice structure}. \hfill \bx}\ed

Suppose $(X,\Delta)$ is a choice structure and $A \subseteq
X$.

\bds \label{def-inverse}{\rm \be \item  $x \in X$ is called
a {\em dynamic element} of $\Delta$ if $\Delta(x) \neq
\emp$; otherwise it is called a {\em static element}. The
set of all dynamic elements of $\Delta$ is denoted by
$dyn(\Delta)$.

\item  $\Delta^{-1}(A) \doteq \{x~|~\emptyset \neq \Delta(x) \subseteq A\}$. $\Delta^{-1}(A)$ is called the
{\em inverse image} of $A$ under $\Delta$. $\Delta^{-1}(y)
\doteq \Delta^{-1}(\{y\})$. Note that $\Delta^{-1}: \sP(X)$
$ \to \sP(X)$.

\item  $\Delta_w^{-1}(A) \doteq \{x~|~\Delta(x) \cap A \neq \emp \}$. $\Delta_w^{-1}(A)$
is called the {\em weak inverse image} of $A$ under
$\Delta$. $\Delta_w^{-1}(y) \doteq \Delta_w^{-1}(\{y\})$.
Note that $\Delta_w^{-1}: \sP(X) \to \sP(X)$. \hfill \bx \ee
} \eds

The examples in Section \ref{sec-exms} may be studied in conjunction
with the
 theory being developed here to help understanding.

\brems \label{rem-inversemap} {\rm Given $X$, $\Delta$, $A$ and
$\{A_j\}$ we have:\be \item  $\Delta^{-1}(\emptyset) = \emptyset =
\Delta_w^{-1}(\emptyset)$.

\item  $\Delta^{-1}(A) \subseteq \Delta_w^{-1}(A) \subseteq dyn(\Delta)$.

\item  $ \Delta^{-1}(\cap_{\j}A_j) = \cap_j \Delta^{-1}(A_j)$ and
$\Delta_w^{-1}(\cup_{\j}A_j) = \cup_j \Delta_w^{-1}(A_j).$

\item $\Delta_w^{-1}(A) = dyn(\Delta)\setminus \Delta^{-1}(A^c)$ and
$\Delta^{-1}(A) = dyn(\Delta)\setminus \Delta_w^{-1}(A^c)$ \hfill
\bx \ee}\erems

\section{Multiplicative and Additive Maps}

\bd \label{def-source} {\rm Suppose  $\mu : \sP(X) \to
\sP(X)$ is a map such that  $\mu(\emp) = \emp$ and  for any
$\{A_j\}$, $\mu(\cap_{\j}A_j) = $ $\cap_j \mu(A_j)$. Then
$\mu$ is said to be a {\em multiplicative map}. $\mu(x)
\doteq \mu(\{x\})$. \hfill \bx }\ed

\brem {\rm If $\Delta$ is a choice set map on $X$  then
$\Delta^{-1}$ is a multiplicative map. \hfill \bx}\erem

\bt \label{th-source}  Suppose $\mu$ is a multiplicative map.\be
\item  If $A_1 \subseteq A_2$ then $\mu(A_1) \subseteq \mu(A_2)$ .

\item  $ \cup_j \mu(A_j)  \subseteq \mu(\cup_{\j}A_j)$ for all $\{A_j\}$.
Equality need not hold.

\item  There exists a unique choice set map $\Delta$ on $X$ such that $\mu =
\Delta^{-1}$. \ee  \et

\pf \be \item   The monotonicity property holds because $ A_1
\subseteq A_2 \Rightarrow A_1 = A_1 \cap A_2 \Rightarrow \mu(A_1) =
\mu(A_1) \cap \mu(A_2) \Rightarrow \mu(A_1) \subseteq \mu(A_2)$.

\item  The inclusion relation follows from the monotonicity property above.
To see that equality need not hold, let $X = \Z$, the set of
integers, and suppose $\Delta(x) = \{\pm x\}$ for all $x \in
\Z$. We then have $\Delta^{-1}([0:\infty)) = \{0\}$ and
similarly $\Delta^{-1}((-\infty:0]) = \{0\}$, but
$\Delta^{-1}(\Z) = \Z$.

\item  Suppose $\mu(X) = B$. By the monotonicity property $A \subseteq X \To
\mu(A)$ $\subseteq B$. Also, for any $x\in B$, there is at least one
$A \subseteq X$ such that $x \in \mu(A)$, namely $A = X$. Suppose
\[ \Delta(x) \doteq \left\{
         \begin{array}{ll}
           \cap \{A ~|~ x \in \mu(A)\}, & \hbox{~if~$x \in B$;} \\
           \emp, & \hbox{~if~$x \notin B$.}
         \end{array}
       \right.\]
Let $x \in B$. We show that then $x \in \mu(\Delta(x))$.
 This proves incidentally that $\Delta(x) \neq \emp$ if and only
if $x \in B$ so that $dyn(\Delta) = B$. By the multiplicative
property $ \mu(\Delta(x)) = \mu(\cap\{A ~|~x \in \mu(A)\})$ $= \cap
\{\mu(A)$ $~|~x \in \mu(A)\}.$ Clearly $x$ is in the set on the
right hand side of the above equality.  So $x \in \mu(\Delta(x)).$

Let $A \subseteq X$. By definition of $\Delta(x)$, $x \in \mu(A)$
$\Rightarrow \emp \neq \Delta(x) \subseteq A \To x \in
\Delta^{-1}(A)$. So $\mu(A) \subseteq \Delta^{-1}(A)$.

Suppose next that $x \in \Delta^{-1}(A)$. Then  $\emp \neq \Delta(x)
\subseteq A$. By the monotonicity property of $\mu$ this implies
that $\mu(\Delta(x)) \subseteq \mu(A)$. We have already seen that
$x$ $\in \mu(\Delta(x))$. So $x \in \mu(A)$. This shows that
$\Delta^{-1}(A) \subseteq \mu(A).$

Combining the last two observations above we see that  $\mu =
\Delta^{-1}$.

To prove the uniqueness of $\Delta$ suppose there are two  choice
set maps $\Delta_1$ and $\Delta_2$ such that $\Delta_1^{-1} $  $=
\mu  = \Delta_2^{-1}$. We have then $dyn(\Delta_1) =
\Delta_1^{-1}(X)$ $= \Delta_2^{-1}(X) = dyn(\Delta_2) = B$, say.

If $x \notin B$, $\Delta_1(x) = \emp = \Delta_2(x)$. Suppose $x \in
B$. Let $\Delta_1(x) = A_1,~ \Delta_2(x) = A_2$. Then $x \in
\Delta_1^{-1}(A_1) \To x \in \Delta_2^{-1}(A_1)$. So $\Delta_2(x)
\subseteq A_1$, or $A_2 \subseteq A_1$. By symmetry $A_1 \subseteq
A_2$. Hence $\Delta_1(x) = \Delta_2(x)$ for all $x \in B$ also, so
that $\Delta_1 = \Delta_2$. \hfill \bx \ee

One can have a characterization of the $\Delta_w^{-1}$ map as of the
$\Delta^{-1}$ map by introducing the notion of an additive map as
below.

 \bd {\rm Suppose $\alpha : \sP(X) \to
\sP(X)$ is a map such that  $\alpha(\emp) = \emp$ and
 for any $\{A_j\}$,  $\alpha(\cup_{\j}A_j) = \cup_j \alpha(A_j)$. Then
 $\alpha$ is called an additive map. $\alpha(x) \doteq
\alpha(\{x\})$. \hfill \bx } \ed

The proof of the next theorem is left to the reader.

 \bt \be \item  Suppose
$\mu$ is a multiplicative map, $\mu(X) = B,$ and for any $A$,
$\alpha_{\mu}(A) \doteq B \setminus \mu(A^c)$. Then $\alpha_{\mu}$
is an additive map.

\item  Suppose $\alpha$ is an additive map, $\alpha(X) = C,$ and for any $A$,
$\mu_{\alpha}(A) \doteq C \setminus \alpha(A^c)$. Then
$\mu_{\alpha}$ is a multiplicative map.

\item  $\mu_{\alpha_{\mu}} = \mu$ and $\alpha_{\mu_{\alpha}} = \alpha$.

\item If $\mu = \Delta^{-1}$ then $\alpha_{\mu} = \Delta_w^{-1}$. If $\alpha =
\Delta_w^{-1}$ then $\mu_{\alpha} = \Delta^{-1}$.
 \hfill
\bx \ee \et

The next theorem gives the properties of additive maps.

\bt Suppose $\alpha$ is an additive map.\be \item  If $A_1 \subseteq
A_2$ then $\alpha(A_1) \subseteq \alpha(A_2)$.

\item  $\alpha(\cap_{\j}A_j) \subseteq \cap_j \alpha(A_j)$ for all $\{A_j\}$.
Equality need not hold.

\item  There exists a unique choice set map $\Delta$ on $X$  such that $\alpha
= \Delta_w^{-1}$.   \ee \et

\pf \be \item   The monotonicity property holds because $A_1
\subseteq A_2 \Rightarrow A_2 = A_1 \cup A_2  \Rightarrow
\alpha(A_2) = \alpha(A_1) \cup \alpha(A_2)\Rightarrow \alpha(A_1)
\subseteq \alpha(A_2).$

\item  The inclusion relation follows from the monotonicity property above.
To see that equality need not hold, let $X = \Z$, the set of
integers, and suppose $\Delta(x) = \{\pm x\}$ for all $x \in
\Z$. We then have $\Delta_w^{-1}([0:\infty)) = \Z$ and
similarly $\Delta_w^{-1}((-\infty:0]) = \Z$, but
$\Delta_w^{-1}(\{ 0 \}) = \{ 0 \}$.

\item   Let $C = \alpha(X)$. By the monotonicity property $A \subseteq X \To
\alpha(A) \subseteq C$. Suppose
\[ \Delta(x) \doteq \left\{
         \begin{array}{ll}
            \{y ~|~ x \in \alpha(y)\}, & \hbox{~if~$x \in C$;} \\
           \emp, & \hbox{~if~$x \notin C$.}
         \end{array}
       \right.\]
 Notice that $C =
\alpha(X) = \cup_{y \in X} \alpha(y)$. So if  $x \in C$ there is at
least one $y \in X$ such that $x \in \alpha(y)$. This proves that
$\Delta(x) \neq \emp$ if and only if $x \in C$ so that $dyn(\Delta)
= C$.

Let $A \subseteq X$. Then $x \in \alpha(A) \Leftrightarrow$
there exists $y \in A$ such that $x \in \alpha(y)
\Leftrightarrow \hbox{ there exists } y \in \Delta(x) \cap A
\Leftrightarrow  \Delta(x) \cap A \neq \emp\Leftrightarrow x
\in \Delta_w^{-1}(A)$. So $\alpha(A) = \Delta_w^{-1}(A)$.

To prove uniqueness, suppose there are two  choice set maps
$\Delta_1$ and $\Delta_2$ such that $(\Delta_1)_w^{-1}$  $ = \alpha
= (\Delta_2)_w^{-1}$. In particular then $dyn(\Delta_1) = $ $
(\Delta_1)_w^{-1}(X) = (\Delta_2)_w^{-1}(X) = dyn(\Delta_2) = C$,
say.

 If $x \notin C$, $\Delta_1(x) = \emp = \Delta_2(x)$.
Suppose $x \in C$.  Then $y \in \Delta_1(x) \To x \in
(\Delta_1)_w^{-1}(y) \To x \in (\Delta_2)_w^{-1}(y) \To \Delta_2(x)
\cap \{ y \} \neq \emp \To y \in \Delta_2(x)$. So $\Delta_1(x)
\subseteq \Delta_2(x)$. By symmetry $\Delta_2(x) \subseteq
\Delta_1(x)$. Hence $\Delta_1(x)$ $ = \Delta_2(x)$ for all $x \in
C$, so that $\Delta_1 = \Delta_2$.\hfill \bx \ee

The results proved so far show that \be
\item There is a one-to-one correspondence between choice set maps
and multiplicative maps. \item
 There is a one-to-one correspondence between
multiplicative maps and additive maps. \item  There is a
one-to-one correspondence between additive maps and choice
set maps. \item The three correspondences commute.\ee

Suppose $(A_n)$ is a sequence of subsets of $X$.  Recall that
$\mathit{lim~sup}~A_n \doteq \cap_{n}(\cup_{k \geq n} A_k)$ and
$\mathit{lim~inf}~A_n \doteq \cup_{n}(\cap_{k \geq n} A_k)$ and that
if both are equal the common value is called $\mathit{lim}~A_n$.
  If
$(A_n\uparrow)$ denotes a monotone increasing sequence of subsets of
$X$ then $lim ~A_n = \cup A_n$. If $(A_n\downarrow)$ denotes a
monotone decreasing sequence of subsets of $X$ then  $lim ~A_n =
\cap A_n$.   $(A_n)$ is said to be a convergent sequence if $\lim
A_n$ exists.

\bd {\rm  A map $\sigma :\sP(X) \to \sP(X)$ is said to be
{\em continuous} if $\sigma(lim~A_n) = lim~\sigma(A_n)$ for
all convergent sequences $(A_n)$.}\ed

\brems {\rm \be \item $\sigma : \sP(X) \to \sP(X)$ is  continuous if
and only if $\sigma(lim~A_n) = lim~\sigma(A_n)$ for all monotone
sequences $(A_n)$. \item The continuity spoken of here is continuity
in the space of sets $\sP(X)$. It needs to be studied how this is
related to the concept of continuity in denotational semantics
\cite{Schmidt}.  \hfill \bx \ee } \erems

 \bt Suppose $\mu$, $\alpha$, and $\Delta$ correspond to one another. The following are equivalent.

\be \item  $\mu$ is continuous.

\item  $\alpha$ is continuous.

\item $\Delta(x)$ is finite for all $x \in X$. \ee \et

 \pf: We have seen earlier that $\mu(X) = \alpha(X) = dyn(\Delta)$. Let this set be denoted by
$B$. We first show that statements (1) and (2) are equivalent.

Since $\mu$ is multiplicative it preserves limits of monotone
decreasing sequences. So $\mu$ is continuous if and only if it
preserves limits of monotone increasing sequences. The situation is
just the other way round  for $\alpha$ because $\mu$ and $\alpha$
are related by the equality $\alpha(A^c) = B \setminus \mu(A)$.

So $\mu \hbox{~is~continuous} \Leftrightarrow
 \mu(\cup A_n) = \cup \mu(A_n), ~\forall~ (A_n\uparrow) \Leftrightarrow
 B \setminus \mu(\cup A_n) = B \setminus \cup \mu(A_n),
~\forall~ (A_n\uparrow) \Leftrightarrow \alpha( \cap A_n^c ) = \cap
\alpha(A_n^c), ~\forall~ (A_n\uparrow) \Leftrightarrow
 \alpha( \cap B_n ) = \cap \alpha(B_n), ~\forall~ (B_n\downarrow) \Leftrightarrow \alpha
 \hbox{~is~continuous}.$

We next show that (1) and (3) are equivalent. For this it is enough
to show that $\Delta(x)$ is finite for all $x$ if and only if
$\Delta^{-1}$ preserves limits of increasing sequences of sets.

 Suppose $\Delta(x)$ is
finite for all $x \in X$ and let $(A_n \uparrow)$. By Theorem
\ref{th-source} we have $\cup (\Delta^{-1}(A_n)) \subseteq
\Delta^{-1}(\cup A_n)$. To prove the opposite inequality suppose $x
\in \Delta^{-1}(\cup A_n)$. Then $\Delta(x) \subseteq \cup A_n$.
Since $\Delta(x)$ is finite and $A_n$'s are monotone, there exists
$m$ such that $\Delta(x) \subseteq A_m$. So $x  \in \Delta^{-1}(A_m)
\subseteq \cup (\Delta^{-1}(A_n))$.

To prove the converse, suppose $\Delta(x)$ is infinite for some $x$,
say $\Delta(x) = \{y_1,y_2,\cdots\}.$ Let $A_n =
\{y_1,y_2,\cdots,y_n\}$. Then, whatever be $n$, $\Delta(x)~
\nsubseteq A_n$ so that $x \notin \Delta^{-1}(A_n)$ and hence $x
\notin \cup\Delta^{-1}(A_n)$. But $x \in \Delta^{-1}(\cup A_n)$. So
$\Delta^{-1}$ is not continuous. \hfill \bx

\section{Convergence}\label{sec-cgce}

\bds {\rm Suppose $\Delta$, $\Delta_1$, and $\Delta_2$ are choice set maps on
$X$.
 \be \item For $A \subseteq X,  \Delta(A) \doteq \cup \{\Delta(x)~|~x \in A\}$, if $A \neq \emp$;
    and $\Delta(\emp) \doteq  \emp$.

\item  For $x \in X$, $(\Delta_2 \circ \Delta_1 )(x) \doteq
\Delta_2(\Delta_1(x))$. $\Delta_2 \circ \Delta_1$ is  a
choice set map on $X$ called the {\em composition} of
$\Delta_2$ with $\Delta_1$.

\item  $\Delta^0(x) \doteq \{ x \}$  so that  $\Delta^0(A) = A$ for any $A
\subseteq X$, and recursively for $k \geq 1$, $\Delta^{k}
\doteq \Delta \circ \Delta^{k-1} \equiv \Delta^{k-1} \circ
\Delta$. \hfill \bx \ee  } \eds \bds \label{def-con}{\rm Let
$\Delta$ be a choice set map on $X$. \be \item $fix(\Delta)
= \{x~|~\Delta(x) = \{ x \}\}$ is called the set of {\em
fixed points} of $\Delta$.
\item  $\mathit{stab}(\Delta) = \{x ~|~\Delta^n(x) \subseteq
\mathit{dyn}(\Delta) \hbox{ for all } n \geq 0 \}$ is called
the set of {\em stable points} of $\Delta$.
\item  $con(\Delta)
= \{x ~|~x \in \mathit{stab}(\Delta),  \Delta^k(x) \subseteq
fix(\Delta)~ \exists~ k \geq 0 \}$ is called the set of {\em
convergent points} of $\Delta$.
\item $con_w(\Delta) = \{x ~|~\Delta^k(x)\cap fix(F) \neq \emp ~
\exists~ k \geq 0 \}$  is said to be the set of {\em weakly
convergent points} of $\Delta$. \hfill \bx \ee  } \eds

 \brems {\rm \label{rem-stab} \be

\item  $fix(\Delta) \subseteq con(\Delta) \subseteq con_w(\Delta) \cap stab(\Delta)$.
\item $\Delta^k(x)\cap fix(F) \subseteq \Delta^{k+1}(x)\cap fix(F)$
for $k \geq 1$.
\item $\Delta^{-1}(stab(\Delta)) \subseteq stab{\Delta}$.  \hfill \bx \ee }\erems

The definitions of convergence and weak convergence given above are
conceptually easy to understand  but  verifying convergence using
these definitions is not convenient in practice. So we give below a
more practical characterization of convergence.

 \bds \label{def-runs} {\rm \be \item  If $y \in \Delta(x)$
 we write $x \mapsto y$ and say that $x$ {\em maps
to} $y$. $\mapsto$ defines a binary relation on $X$.

\item  For $n \geq 1$,  a finite sequence $(x_0,x_1,\cdots,$ $x_n)$
of elements of $X$ is called a {\em run} of {\em length} $n$
{\em starting} at $x_0$ and {\em ending} at $x_n$ if $ x_0
\mapsto x_1 \mapsto \cdots \mapsto x_n$.  In this case $x_1
\in \Delta(x_0),~ x_2 \in \Delta(x_1),~ \cdots~, x_n \in
\Delta(x_{n-1})$. Also $x_n \in \Delta^n(x_0)$.
\item  If  $(x_0,x_1,\cdots,x_n)$ is a run we write $x_0$ $ \yields x_n$ and
say that  $x_n$ is {\em reachable} from $x_0$. It may be
observed that $\yields$ is the transitive closure of
$\mapsto$.
\item If $(x_0,x_1,\ldots,x_n)$ is a
run and $0 < m < n$ then $(x_0,x_1,\ldots,x_m)$ is also a
run. In such a case, we say that $(x_0,x_1,\ldots,x_n)$ is
an {\em extension} of $(x_0,x_1,\ldots,x_m)$.
\item  A run is said to be {\em aborted} if it ends in a state that is not in
$dyn(\Delta)$; that is, if it has no extension.
\item   A run is said to be {\em terminal} if it ends in a fixed point of $\Delta$.
 If $(x_0,x_1,\cdots,x_n)$ is a terminal run and $m$ is the
least positive integer such that $x_m \in fix(F)$, then $x_m
= x_{m+1} = \cdots = x_n$. \hfill \bx \ee } \eds

\bt Suppose $\Delta$ is a choice set map on $X$ and $x \in X$. \be
\item $x \in con(\Delta)$ if and only if \be \item   there are runs at $x$; \item every run at $x$  can be extended; \item there
exists $k \geq 1$ such
 that every run starting at
$x$ of length  $k$ or more is a terminal run.\ee
\item  $x \in con_w(\Delta)$ if and only if there exists a run starting at
$x$ that is terminal. \ee \et

\pf \be \item  Suppose that $x \in con(\Delta)$ so that $\Delta^n(x)
\subseteq \mathit{dyn}(\Delta)$ for all $n \geq 0$, and $\Delta^k(x)
\subseteq
fix(F)$ for some $k \geq 1$.  We need to prove $(a),(b),$ and $(c)$.\\

Since $x \in dyn(\Delta)$, $\Delta(x)$ is not empty. Let $y \in \Delta(x)$. Then $(x,y)$ is a run. So there are runs at $x$. This argument can be repeated with the last element of the run replacing $x$ above. This shows that any run at $x$ can be extended.\\

Consider any run $(x, x_1,x_2,\cdots,x_k)$ of length $k$. Then $x_k
\in \Delta^k(x) \subseteq fix(F)$. Hence it is a terminal run.\\

Conversely assume that $x$ satisfies $(a),(b)$ and $(c)$. Since
there are runs at $x$, $x \in \mathit{dyn}(\Delta).$ Since every run
at $x$ can be extended it follows that $\Delta^n(x) \subseteq
\mathit{dyn}(\Delta)$ for every $n \geq 0$. So $x$ is stable. Let
$k$ be given by $(c)$.  We need to prove  that $\Delta^k(x)
\subseteq fix(F)$. Let $x_k \in \Delta^k(x)$. Since $k > 1$,
$\Delta^k(x) = \Delta(\Delta^{k-1} (x))$.  So there exists $x_{k-1}
\in \Delta^{k-1}(x)$ such that $x_k \in \Delta(x_{k-1})$. Continuing
in this way we can construct a run $(x,x_1,\cdots,x_k)$. Since this
run has length $k$ it is terminal. So $x_k \in fix(F)$. Since $x_k$
was chosen to be an arbitrary element in $\Delta^k(x)$ it follows
that $\Delta^k(x) \subseteq fix(F)$.

\item  Suppose $x \in con_w(\Delta)$. Then there exists $k \geq 1$ such that
$\Delta^k(x) \cap fix(F) \neq \emp$. So there exists an element $x_k
\in \Delta^k(x) \cap fix(F)$. Since $x_k \in \Delta^k(x) =
\Delta(\Delta^{k-1}(x))$ and  $k
> 1$, there exists $x_{k-1}
\in \Delta^{k-1}(x)$ such that $x_k \in \Delta(x_{k-1})$. Continuing
inductively we get a sequence $x_i, 1 \leq i \leq k$ such that $x
\mapsto x_1
\mapsto \cdots \mapsto x_k \in fix(F)$. Its length is $k$.\\

Conversely assume that there exists a terminal run starting at $x$
of length $k$. Then there exist $x_i$,   $1 \leq i \leq k$,  such
that
 $x  \mapsto x_1 \mapsto \cdots \mapsto x_k \in fix(F)$.
Then  $x_k \in \Delta^k(x)$.  Hence $\Delta^k(x) \cap fix(F) \neq
\emp$.
 \hfill \bx\ee

Let $\Delta$ be a choice set map on $X$.  We had observed in Remarks
\ref{rem-stab} that the sequence of sets $\Delta^k(x) \cap fix(F)$
is monotonically increasing.

\bd \label{def-H} {\rm Let $(X, \Delta)$ be a choice structure.
  For any $x \in X$ define  $\Di(x) =  \cup~(\Delta^k(x) \cap
fix(F)).$  The choice set map $\Di$ is called the {\em limit
map} of $\Delta$.  Elements of $\Di(x)$ are called the {\em
limit points} of $\Delta$  at  $x$.  \hfill \bx }\ed

\brems {\rm \be \item  $fix(\Di) = \mathit{fix}(\Delta)$.

\item  $\Di(x) = \{ y: x \yields y, y \in fix(F) \}$. These are the points of $fix(F)$
that are reachable from $x$.

\item $x \in con(\Delta)$ if and only if every run at $x$ when sufficiently extended ends up in $fix(\Delta)$. The set of all such reachable points of $fix(\Delta)$ is precisely $\Di(x)$.

\item  $x \in con_w(\Delta)$ if and only if $\Di(x) \neq \emp$ so that $dyn(\Di) =
con_w(\Delta)$.

\item  Suppose $\Delta(x)$ is finite for all
$x$. If $x  \in con(\Delta)$ then there exists $k \geq 1$ such that
$\Delta^k(x) \subseteq fix(F)$. In such a case $\Delta^k(x) =
\Di(x)$. In particular $\Di(x)$ is finite.
 So it is impossible to have a
convergent choice structure with $\Delta(x)$ finite and $\Di(x)$
infinite for
 $x \in X$. It is this fact that Dijkstra is pointing out when he says
\cite{Dijk76}  that there can not exist a program that says ``set
$x$ to any positive integer". Example \ref{exm-unbdd} illustrates
this point further. \hfill \bx \ee }\erems

\bd {\rm  For any  $A \subseteq X$  the set $bas(\Delta, A)=
\{ x \in con(\Delta) ~|~ \Delta^{\infty}(x) \subseteq A \}$
is called the {\em basin} of $A$ with respect to $\Delta$ .
\hfill \bx }\ed

\brems \label{rem-basin} {\rm \be \item Recalling Definition
\ref{def-inverse} we see that $bas(\Delta, A) = con(\Delta) \cap
(\Di)^{-1}(A)$ for all $A \subseteq X$.

\item It is not true that $bas(\Delta, A) = (\Di)^{-1}(A)$ for all $A \subseteq X$ if $con(\Delta) \neq con_w(\Delta)$.
 For let $x \in con_w(\Delta) \setminus con(\Delta)$ and  take $A = fix(\Delta)$. Then $\Di(x) \subseteq A$ so that $x \in (\Di)^{-1}(A)$ but $x \notin bas(\Delta,A)$.
\hfill \bx\ee }\erems

\section{Examples}\label{sec-exms}

\bxm {\rm \label{exm-abort} Suppose $X$ is any set and  $\Delta(x)$
$ = \emp$ for all $x$. Then $dyn(\Delta) =  fix(\Delta) = \emp$,
$\Delta_w^{-1}(A) = \Delta^{-1}(A) = \emp$ for all $A \subseteq X$,
and
 $\Delta^k(x) = \emp$ for all $x \in X$. No state maps to any state nor yields any state.
 There are no runs, no
stable points, no convergent points and no weakly convergent points.
So $stab(\Delta) = con(\Delta)  = con_w(\Delta) = \emp$. Also
$\Delta^{\infty}(x) = \emp$, $x \in X$ and $bas(\Delta, A) = \emp$
for every $A \subseteq X$.  $\Delta$ may be identified with the
$abort$ command. \hfill \bx}\exm

\bxm{\rm \label{exm-skip} Suppose $X$ is any set and
$\Delta(x) = \{x\}$ for all $x \in X$. Then $dyn(\Delta) =
fix(\Delta) = X,$ $\Delta_w^{-1}(A) = \Delta^{-1}(A) = A$
for all $A \in \sP(X)$, and $\Delta^k(x) = \{x\}$ for all
$x$. Every element maps only to itself and yields only
itself. Every run is of the form $(x, x, \cdots , x)$ and is
terminal. Every element yields only itself. So $stab(\Delta)
= con(\Delta) = con_w(\Delta) = X$. $\Delta^{\infty}(x) =
\{x\}$ for all $x \in X$. $bas(\Delta, A) = A$ for all $A
\in \sP(X)$. This structure may be identified with $skip$,
because leaves everything unchanged. \hfill \bx}\exm

\bxm {\rm \label{exm-unbdd}  Suppose $X$ is any infinite set and
$\Delta(x) = X$ for all $x$. Then , $dyn(\Delta) = stab(\Delta) =
X$, $fix(\Delta) = \emp$. $\Delta^{-1}(X) = X$ and $\Delta^{-1}(A) =
\emp,$ if $A \neq X$. $\Delta_w^{-1}(A) = X$ if $\emp \neq A
\subseteq X$. $\Delta^{k}(x) = X$ for all $x \in X$ and $k \geq 1$.
Every element maps to every other element and yields every other
element. Any finite sequence of elements of $X$ is a run and no run
is terminal. There are no convergent points or weakly convergent
points, so that $con(\Delta) = \emp = con_w(\Delta)$.
$\Delta^{\infty}(x) = \emp$ for all $x \in X$. $bas(\Delta, A) =
\emp$ for every $A \in \sP(X)$. \hfill \bx}\exm

\bxm {\rm \label{exm-det} Suppose $F : X \to X$ is a map and
$\Delta(x) = \{F(x)\}$ for all $x \in X$. We call $\Delta$ a
deterministic map. In this case $dyn(\Delta) = X$,
$stab(\Delta) = X$, $fix(\Delta) = fix(F)$, and all the
definitions we have given above reduce to the corresponding
definitions for the deterministic flow $(X,F)$ as given in
\cite{KVbook}. We have $\Delta^k(x) = \{F^k(x)\}$ for all
$x$ and $k$. $con(\Delta)  = con_w(\Delta) = con(F)$.
$\Delta_w^{-1}(A) = \Delta^{-1}(A) = F^{-1}(A)$ for all $A
\in \sP(X)$.  Further $\Delta^{\infty}(x) =
\{F^{\infty}(x)\}$ for all $x \in con(F),$ and
$bas(\Delta,A) = \cup_{k\geq 0} F^{-k}(A \cap fix(F))$.
\hfill \bx}\exm

\bxm {\rm \label{exm-N}Let $X = \N$ and suppose $\Delta$ is a choice
set map on $\N$ defined by \[\Delta(x) = \left\{
  \begin{array}{ll}
    \{0\}, & \hbox{~if~}x = 0; \\
    \{x-1,x+1\}, & \hbox{~if~}x > 0.
  \end{array}
\right.\]

Then $dyn(\Delta) = stab(\Delta) =  \N$ and $fix(\Delta) = \{0\}$.
If $x
> 0$, $(x,x-1,x-2,\cdots,0)$ is a terminal run of length $x$. It
follows that every state is weakly convergent.

 It may be noted that for any $x$ and $n > 0$, a run of the form
$(x,x+1,x+2,\cdots, x+n, x+n-1, \cdots, x,x-1,x-2,\cdots,0)$ is also
terminal with length $x + 2n$. So there exist arbitrarily long
terminal runs at any $x$. At the same time for any $x$,
$(x,x+1,x+2,\cdots,x+n)$ is a nonterminal run for every $n$. So
there also exist arbitrarily long nonterminal runs starting at every
$x
> 0$. It follows that no state is   convergent except $0$.

For this example $con(\Delta) = fix(\Delta) = \{0\}$ and
$con_w(\Delta) = X$. $\Delta^{\infty}(x) = \{0\}$ for all $x \in
\N$. $bas(\Delta, A) = \{0\} \Leftrightarrow 0 \in A$.  \hfill
\bx}\exm

\bxm {\rm \label{exm-Z} In this example we show that the sets
$dyn(\Delta),  stab(\Delta), con(\Delta), con_w(\Delta)$ can all be
different. Let $X = \Z$ and suppose $\Delta$ is defined by
\[\Delta(x) = \left\{
  \begin{array}{ll}
    \{x-2, x+2\}, & \hbox{ if $x \geq  0, x \neq 2$;} \\
    \{ 2 \}, & \hbox{ if } x = 2,\\ \emp, & \hbox{~if~} x < 0.
  \end{array}
\right.\]

Then $dyn(\Delta) =  \N$, $stab(\Delta) = 2\N \setminus \{
0 \}$, and $fix(\Delta) = \{2\}$. If $x > 0$ is an odd
number, $\Delta^k(x)$ contains only odd numbers and hence
$\Delta^{k}(x) \cap fix(\Delta) = \emp$ for every $k \geq
0$.

If $x = 0$, $(0, 2)$ is a terminal run of length 1. If $x > 0$ is
even, $(x,x-2,x-4,\cdots,0)$ is a terminal run of length $x/2$. It
may be noted that $(0, -2)$ is an aborted run.  For any $x > 2$, $x$
even, and $n > 0$, a run of the form $(x,x+2,x+4,\cdots, x+2n,
x+2n-2, \cdots, x,x-2,x-4,\cdots,2)$ is a terminal with length $2n +
(x/2) - 1$. So there exist arbitrarily long terminal runs at any
even $x, x > 2$. At the same time for any such $x$,
$(x,x+2,\cdots,x+2n)$ is a nonterminal run for every $n$. So there
also exist arbitrarily long nonterminal runs starting at every even
$x
> 2$.

It follows that $con_w(\Delta) = 2\N$ and  $con(\Delta) =  \{2\}$.
$\Delta^{\infty}(x) = \{2\}$ for all $x \in 2\N$. $bas(\Delta, A) =
\{2\} \Leftrightarrow 2 \in A$ and $bas(\Delta, A) = \emp
\Leftrightarrow 2 \notin A$.  \hfill \bx}\exm

 \bxm {\rm \label{exm-N^2}In the above example we saw that there exist $x \in X$
such that there could be terminal runs of arbitrary length starting
at $x$. However all of the runs end up in the same final state. The
present example \cite{Dijk76} is one where there are terminal runs
of arbitrary length that start at the same state but end up at
different states.

Let $X = \N \times \{0,1\}$. Define $\Delta$ by
\[\Delta(x,y) = \left\{
  \begin{array}{ll}
    \{(x,0),(x+1,1)\}, & \hbox{~if~$y =1$;} \\
    \{(x,0)\}, & \hbox{~if~$y = 0$.}
  \end{array}
\right.\]

It is left to the reader to check that $fix(\Delta)  = con(\Delta) =
\N \times \{0\}$, $stab(\Delta)
 = con_w(\Delta) = X$, $\Delta^{\infty}(x,0) = \{(x,0)\}$ and
$\Delta^{\infty}(x,1) = \{ (x+n, 0) ~|~ n \geq 0 \}$.} \hfill \bx
\exm

\section{Dijkstra's $IF$ and $DO$ constructs} \label{sec-dijk}

After describing the concept of a state and introducing the state
space (which we have called $X$) Dijkstra \cite{Dijk76}(p.15)
introduces the notion of a nondeterministic machine. He says that
``activation (of such a machine) in a given initial state will give
rise to one out of a class of possible happenings, the initial state
only fixing the class as a whole". We have interpreted this
statement to mean that for every $x \in X$ we are given a set
$\Delta(x) \subseteq X$ such that if $x$ is the initial state, then
$\Delta(x)$ is the set of all possible happenings when the
nondeterministic machine is invoked once. Thus a choice structure
$(X,\Delta)$ is our model for
a nondeterministic machine.\\

 However, even  after almost
defining a choice structure, Dijkstra does not formalize
nondeterminism in this way. He says that ``the design of
such a system is a goal-directed activity, in other words
that we want to achieve something with that system." What we
want to achieve is a ``post-condition". That is to say after
the machine is invoked we want to insure that the resulting
state belongs to a certain set $A \subseteq X$. He then says
that ``we should like to know .... the set of (all) initial
states such that activation will certainly result in a
properly terminating happening leaving the system in the
final state satisfying the post-condition". In our notation
this is $\Delta^{-1}(A)$. This  set he calls the ``weakest
pre-condition" and denotes it by $wp~(S,A)$, where $S$ is
his notation for the mechanism. Without giving a definition
of $S$ directly he wants to characterize it by the map $A
\mapsto wp~(S,A)$. He shows that, as we have done in
Section~\ref{sec-defns}, that this map is multiplicative in
$A$. So, for Dijkstra, every nondeterministic mechanism is
given by a multiplicative map. In our notation we shall
henceforth take  $wp~(\Delta,A)$ to be the same as
$\Delta^{-1}(A)$.

We need now to connect the theory of nondeterminism developed so far
using $\Delta$ to the theory that may be developed using
$\Delta^{-1}$. Before doing that, we shall define the structures
$IF$ and $DO$ directly in terms of choice set maps and derive the
two main theorems about them to show how
simple the definitions and proofs are in our approach.\\

A patch on $X$ is a pair $(D,F)$ where $D \subseteq X$ and $F : D
\to X$ \cite{KVbook}. A patch $(D,F)$ can be interpreted to be a
guarded command. Its action is first to check if a given state $x$
is in $D$. If it is, $x$ is changed to $F(x)$. If it is not, then no
action is taken\footnote{In Dijkstra's definition of a guarded
command $(D,F)$ the map $F$ is taken to be a global map, for certain
technical reasons which do not concern us here.}.

\bds {\rm \be \item A quilt $Q$ is a collection of patches: $Q = \{
(D_1,F_1),(D_2,F_2),\cdots,(D_k,F_k) \}$.

\item Given
a quilt $Q$ let   $D = \cup_{1 \leq i \leq k} D_i$ and
define the choice set map $\Delta_Q$  by \beqa \Delta_Q(x)
&=& \left\{
  \begin{array}{ll}
    \{F_i(x)~|~x \in D_i \hbox{~for~some~} i\} & \hbox{~if~}x \in D; \\
    \{x\}, & \hbox{~if~} x \notin D
  \end{array}
\right. \eeqa \hfill \bx \ee }\eds

By definition, $\Delta_Q(x) \neq \emp$ for all $x \in X$. So
$dyn(\Delta_Q) = X$. What about $fix(\Delta_Q)$? Clearly $D^c
\subseteq fix(\Delta_Q)$. There could be points of $D$ also in
$fix(\Delta_Q)$. Let $E = \{x \in D~|~x \in D_i \To F_i(x) = x \}$.
Then $E \subseteq fix(\Delta_Q)$ and in fact $fix(\Delta_Q) = D^c
\cup E$. It is to be noted that the set $E$ is not mentioned
explicitly by Dijkstra.

\bd {\rm Let $Q$ be a quilt as above let
 $D = \cup_{1 \leq i \leq k} D_i$. The choice structure $\Delta_{IF}$ is defined by \beqa
\Delta_{IF}(x) &=& \left\{
  \begin{array}{ll}
    \Delta_Q(x) & \hbox{~if~}x \in D; \\
    \emp, & \hbox{~if~}x \notin D
  \end{array}
\right.\eeqa \hfill \bx }\ed

We then have   $dyn(\Delta_{IF}) = D$ and $fix(\Delta_{IF}) = E$.

 The ``basic theorem for the alternative construct" takes
 the following form. The proof follows immediately
 from the definitions of $\Delta_Q$ and $\Delta_{IF}$.

\bt  Let $A, B \subseteq X$ be such that  $A \subseteq D$, and
$F_j(A \cap D_j) \subseteq B$ for all $j$. Then $\Delta_{IF}(A)
\subseteq B$. \et

Next, let us consider the repetitive construct $DO$. It seems
natural to define it by either $\Di_Q$ or $\Di_{IF}$. However,
Dijkstra does neither, for two reasons. The first is that he does
not want the points that are weakly convergent for $\Delta_Q$, but
not convergent, in the domain of $DO$. Secondly, he does not
consider the computation terminated unless the state enters $D^c$.
This means that if the state finds itself in the set $E$ then, even
though it is a fixed point, for both $\Delta_Q$ and $\Delta_{IF}$,
the computation is not considered to terminate: The points of $E$
should be considered to be the points where the computation
``hangs". To construct a properly terminating program guaranteeing
an outcome in $D^c$ we need therefore to take away from
$dyn(\Delta_Q)$ all the points that are weakly convergent but not
convergent, and also all those points that end up in $E$. This means
that we need to restrict ourselves to the set $bas(\Delta_Q,D^c)$.
By Remark~\ref{rem-basin} this is the set $con(\Delta_Q) \cap
(\Di_Q)^{-1}(D^c)$. We have then the following definition.

\bd \label{def-DO}{\rm Let a quilt $Q$ be given as above. Then the
choice structure $\Delta_{DO}$ is defined by
 \beqa
\Delta_{DO}(x) &=& \left\{
  \begin{array}{ll}
    \Di_Q(x), & \hbox{~if~$x \in bas(\Delta_Q, D^c)$}, \\
    \emp, & \hbox{~otherwise.}
  \end{array}
\right. \eeqa } \hfill \bx \ed

Clearly  $dyn(\Delta_{DO}) = bas(\Delta_Q, D^c)$ and
$fix(\Delta_{DO}) = D^c$. Also $\Delta_{DO}(X) \subseteq D^c$.

The ``fundamental invariance theorem for loops" takes the
following form. \bt  Let $V \subseteq X$ be such that
$\Delta_{IF}(V \cap D) \subseteq V$.  Then $\Delta_{DO}(V
\cap con(\Delta_Q)) \subseteq V \cap D^c$.\et

\pf Since $\Delta_{IF} = \Delta_Q$ on $D$ we are given that
$\Delta_Q(V \cap D) \subseteq V$.

 Let $x \in V \cap con(\Delta_Q))$. If $x \in D^c$ there is nothing to prove. So let $x \in D$.  Since
$x \in con(\Delta_Q)$ there exists $k \geq 0$ such that $\Delta^k(x)
= \Di_Q(x)$. Let $y \in \Delta^k(x)$. Then $y \in D^c$ and there
exists a run $x = x_0,x_1,\cdots,x_k = y$. Let $j$ be the least
integer such that $x_j \in D^c$. Then $x_j = y$ and
$x_0,x_1,\cdots,x_{j-1} \in D$.  So we have successively
\beqa x_0 \in V \cap D &\To& x_1 = \Delta_Q(x_0) \in V \cap D \\
&\To& x_2 = \Delta_Q(x_1) \in V \cap D \\
&\vdots&\\
&\To& x_{j-1} = \Delta_Q(x_{j-2}) \in V \cap D \\
&\To& x_j = \Delta_Q(x_{j-1}) \in V \cap D^c. \eeqa

The theorem is proved. \hfill \bx

We have thus seen that using the formalism of choice set
maps it is very easy to understand the structures $IF$ and
$DO$. We now need to prove that our definitions coincide
with Dijkstra's.

Let us consider $IF$ first, and let us consider the special case
when there is only one patch $(D,F)$. In this case $wp~(F,A)$ is
described on p.17 of \cite{Dijk76} by the following sentence
(notation changed): ``If the initial state satisfies $wp~(F,A)$, the
mechanism is certain to establish eventually the  truth of $A$".
This means that if $x \in wp~(F,A)$ then $F(x) \in A$. Or $wp~(F,A)
= F^{-1}(A)$.

Consider next the case of a general quilt $Q$ as above. Then
\beqa \Delta_{IF}^{-1}(A) &=& \{x~|~\emp \neq \Delta_{IF}(x) \subseteq A\}\\
            &=& \{x~ \in D~|~\Delta_{IF}(x) \subseteq A\}\\
&=& \{ x \in D~|~x \in D_i \To F_i(x) \in A\}\\
&=& D \cap  \{x~|~x \in D_i \To x \in F_i^{-1}(A) \}\\
&=& D \cap  \{x~|~x \in D_i \To x \in wp~(F,A) \}\\
\eeqa

But this is exactly the definition of $wp~(IF,A)$ on   p.34 of
\cite{Dijk76}. So $wp~(IF,A) = \Delta_{IF}^{-1}(A)$ for all $A
\subseteq X$.

For the $DO$ construct also we need to show that
$\Delta_{DO}^{-1}(A) = wp~(DO,A)$ for all $A \subseteq X$. This
takes some hard work. By Definition~\ref{def-DO} we see that
$\Delta_{DO}^{-1}(A) = bas(\Delta_Q, A \cap D^c)$. So we need to
show that $ bas(\Delta_Q, A \cap D^c) = wp~(DO,A)$. For this
purpose, we need to first characterize the set $bas(\Delta,A)$ in
terms of iterates of $\Delta^{-1}$ for any choice set map $\Delta$
and for any $A \subseteq X$.

\bd {\rm Given a choice set map $\Delta$  and $A \subseteq X$,
$\Delta^{-k}(A) \doteq (\Delta^{-1})^k(A)$ for $k \geq 1$ and
$(\Delta^{-1})^0 \doteq \Delta^0$. \hfill \bx}\ed It is natural to
ask ourselves at this stage how $(\Delta^{-1})^k$ is related to
$(\Delta^k)^{-1}$. First of all we note that they need not be equal.
\bxm{\rm  Let $X = \{a,b,c\}$, and let $\Delta(a) = \{ b,c \}, ~
\Delta(b) = \emp, ~ \Delta(c) = \{c\}$. Then $\Delta^2(a) = \{c\}$,
so that $a \in (\Delta^2)^{-1}(c)$. But $(\Delta^{-1})^2(c) =
\Delta^{-1}(c) = \{c\}$. }\hfill \bx \exm We have the following
result. \bl Let $\Delta$ be a choice set map on $X$, $A \subseteq
X$, and $k>0$. Then
 \be \item  $(\Delta^{-1})^k(A)
\subseteq (\Delta^k)^{-1}(A)$; \item $(\Delta^k)^{-1}(A) \cap
\mathit{stab}(\Delta) \subseteq (\Delta^{-1})^k(A).$ \ee \el

\pf We prove the theorem by induction on $k$. \be
\item For  $k=1$ equality holds. Assume the result for $k$.
\beqa x \in (\Delta^{-1})^{k+1}(A) &\To& \emp \neq \Delta(x)
\subseteq (\Delta^{-1})^{k}(A)\\ &\To&  \emp \neq  \Delta(x)
\subseteq (\Delta^k)^{-1}(A)\\ &\To&  \emp \neq \Delta^{k+1}(x)
\subseteq A \\ &\To& x \in (\Delta^{k+1})^{-1}(A)\\ \eeqa

Hence  $(\Delta^{-1})^{k+1}(A) \subseteq (\Delta^{k+1})^{-1}(A)$.

\item  For $k=1$ the relation holds. Assume that $(\Delta^k)^{-1}(A) \cap \mathit{stab}(\Delta) \subseteq
(\Delta^{-1})^k(A)$. Then

\beqa x \in (\Delta^{k+1})^{-1}(A) \cap \mathit{stab}(\Delta) & \To
& x \in \mathit{stab}(\Delta) \hbox{ and } \emp \neq
\Delta^{k}(\Delta(x)) = \Delta^{k+1}(x) \subseteq A,  \\
& \To & x \in \mathit{stab}(\Delta) \hbox{ and } \emp \neq
\Delta^{k}(y) \subseteq A \hbox{ for every $y \in \Delta(x)$}\\
& \To & x \in \mathit{stab}(\Delta) \hbox{ and } y \in (\Delta^k)^{-1}(A) \hbox{ for every $y \in \Delta(x)$}\\
& \To & x \in \mathit{stab}(\Delta) \hbox{ and } y \in (\Delta^k)^{-1}(A) \cap stab(\Delta) \hbox{ for every $y \in \Delta(x)$}\\
& \To & x \in \mathit{stab}(\Delta) \hbox{ and } y \in (\Delta^{-1})^k(A) \hbox{ for every $y \in \Delta(x)$,} \\
 & \To & \emp \neq \Delta(x) \subseteq  (\Delta^{-1})^k(A)  \\
& \To & x \in (\Delta^{-1})^{(k+1)}(A) \eeqa

This proves that $(\Delta^{k+1})^{-1}(A) \cap \mathit{stab}(\Delta)
\subseteq (\Delta^{-1})^{k+1}(A)$. \bx \ee

\brem \label{rem-stabinv}{\rm It follows from Remark~\ref{rem-stab}
that  $\Delta^{-k}(stab(\Delta)) \subseteq \mathit{stab}(\Delta)$,
for $k \geq 0$. In particular $\Delta^{-k}(fix(\Delta)) \subseteq
\mathit{stab}(\Delta)$, for $k \geq 0$.\hfill \bx } \erem

\bt \label{th-basin} For any $A \subseteq X$, $bas(\Delta,A) =
\cup_{k \geq 0} \Delta^{-k}( A \cap fix(\Delta) )$. \et

\pf It is enough to consider the case $A \cap fix(\Delta) \neq
\emp$.

Suppose $x \in bas(\Delta, A)$. Then $ x \in con(\Delta) \subseteq
\mathit{stab}(\Delta)$, and there exists $k \geq 0$ such that $\emp
\neq \Delta^{\infty}(x) = \Delta^{k}(x) \subseteq A \cap
fix(\Delta)$. This implies that $x \in (\Delta^k)^{-1}( A \cap
fix(\Delta) ) \cap \mathit{stab}(\Delta) \subseteq (\Delta^{-1})^k(A
\cap fix(\Delta))$ and hence $x \in  \Delta^{-k}(A \cap
fix(\Delta)$.

 So $bas(\Delta,A) \subseteq \cup_{k \geq 0}
\Delta^{-k}( A \cap fix(\Delta) ).$

 Conversely, suppose   $x \in
\Delta^{-k}( A \cap fix(\Delta) )$ for some $k \geq 0$. By the
remark \ref{rem-stabinv}, $x \in \mathit{stab}(\Delta)$ also. Since
$(\Delta^{-1})^k(A \cap fix(\Delta)) \subseteq (\Delta^k)^{-1}(A
\cap fix(\Delta))$, we have $\Delta^{k}(x) \subseteq A \cap
fix(\Delta)$.  Then $x \in con(\Delta)$ and $\Delta^{\infty}(x) =
\Delta^{k}(x) \subseteq A$ so that $x \in bas(\Delta,A)$.  So
$\cup_{k \geq 0} \Delta^{-k}( A \cap fix(\Delta) )$ $\subseteq
bas(\Delta,A).$

This proves that $bas(\Delta,A) = \cup_{k \geq 0} \Delta^{-k}( A
\cap fix(\Delta) ).$ \hfill \bx

To complete the connection to Dijkstra's $wp$ formalism we need to
connect the map $\Delta_{DO}^{-1}$ with the iterates of
$\Delta_{IF}^{-1}$.

\bl Let $A \subseteq X$.  Define $H_0(A) = A \cap D^c $, and for $k
> 0$, $H_{k+1}(A) = wp~(IF,H_k(A)) \cup H_0(A)$. Then
$\Delta_Q^{-k}(A \cap D^c) = H_k(A) $ for all $k \geq 0$.  \hfill
\bx \el

\pf  Note first that if $x \in H_0(A)$, then $\Delta_Q(x) = \{x\}
\subseteq A \cap D^c$ so that $H_0(A) \subseteq \Delta_Q^{-1}(A \cap
D^c)$. By the monotonicity property of multiplicative maps we  have
$H_0(A) \subseteq \Delta_Q^{-k}(A \cap D^c)$ for all $k > 0$.

For $k = 0$ we have $H_0(A) = A \cap D^c = (\Delta_{IF}^{-1})^0(A
\cap D^c)$. Assume that $\Delta_Q^{-k}(A \cap D^c) = H_k(A)$ for
some $k$. Then \beqa x \in H_{k+1}(A)
 & \Leftrightarrow & x \in wp~(IF,H_k(A))
  \hbox{ or } x \in H_0(A)\\
 & \Leftrightarrow & x \in (\Delta_{IF})^{-1}(H_k(A))
  \hbox{ or } x \in H_0(A)\\
& \Leftrightarrow & x \in D \hbox{ and } \Delta_{IF}(x)
\subseteq H_k(A)  \hbox{ or } x \in H_0(A)\\
& \Leftrightarrow & x \in D \hbox{ and } \Delta_Q(x) \subseteq
\Delta_Q^{-k}(A \cap D^c)  \hbox{ or } x \in H_0(A)\\
 & \Leftrightarrow & x \in D \hbox{ and } x \in
\Delta_Q^{-1}(\Delta_Q^{-k}(A \cap D^c)) \hbox{ or } x \in H_0(A)\\
& \Leftrightarrow & x \in \Delta^{-(k+1)}(A \cap D^c).  \eeqa

This proves the lemma. \bx

\bt $\Delta_{DO}^{-1}(A) = wp~(DO,A)$ for all $A \subseteq X$. \et

\pf For the proof we only need to collect our earlier results and
see the definition of $wp~(DO,A)$ on p.35 of \cite{Dijk76}.

\beqa \Delta_{DO}^{-1}(A) &=& bas(\Delta_Q,A \cap D^c)\\
&=& \cup_{k \geq 0}\Delta_Q^{-k}(A \cap D^c)\\
&=& \cup_{k \geq 0}H_k(A)\\
&=& wp~(DO,A) \eeqa This proves the theorem. \bx

\section{Concluding Remarks} Dijkstra \cite{Dijk76} introduces the
notion of a nondeterministic mechanism acting on a state
space $X$ but does not define the notion. Rather he says
that such a mechanism induces a set action that we have
denoted by $\mu$ and that the action characterizes the
mechanism. We have shown that $\mu$ is determined by a
choice set map and that the backward acting $\mu$ is
equivalent to the forward acting $\Delta$. Thus this article
presents an alternative approach to the understanding of
Dijkstra's formalism. We have also shown that there is a
third way and equivalent way of defining nondeterminism that
is dual to that of Dijkstra, in  terms of additive maps.

Our approach also suggests there is a weak convergence
related to additive maps that could operate in
nondeterministic mechanisms. In subsequent articles we shall
choose the choice set map as our primary way of modeling
nondeterminism and present an exposition of the design of
algorithms as suggested by Dijkstra, and also the standard
concepts of computability, complexity, witness certificates
and other such ideas studied in a standard course in the
theory of computation \cite{Papa}. It turns out that weak
inverses of choice set maps have an important role to play.

\bibliographystyle{acmtrans}
\bibliography{kv-bibliography-may2009}

\begin{received}
...
\end{received}

\end{document}